\documentstyle[12pt]{article}
\begin{document}
\large

\begin{center}{\large\bf ASPECTS OF THE THEORY OF
DEEP INELASTIC SCATTERING}\end{center}
\begin{center} F.M. Lev\end{center}
\begin{center}{\it Laboratory of Nuclear Problems,
Joint Institute for Nuclear Research, Dubna, Moscow region 141980,
Russia}\end{center}

\begin{abstract}
The (electromagnetic or weak) current operator responsible for
deep inelastic scattering (DIS) should be local and satisfy the
well-known commutation relations with the representation operators
of the Poincare group. The problem whether these conditions are
compatible with the factorization theorem and operator product
expansion is investigated in detail. We argue that the current
operator contains a nontrivial nonperturbative part which contributes
to DIS even in the Bjorken limit. Nevertheless there exists a
possibility that many results of the standard theory remain.
\end{abstract}

\section{The statement of the problem}
\label{S1}

 The recent discovery of diffractive deep inelastic scattering (DIS)
at HERA \cite{dif,dif1} poses the problem whether DIS processes with
small values of the Bjorken variable $x_{Bj}$ can be described in the
framework of perturbative QCD (see e.g. ref.
\cite{kaid} and references therein). At the same time the
applicability of perturbative QCD in the Bjorken limit (i.e. in the case
when the momentum transfer $q$ is such that $Q=|q^2|^{1/2}$ is large and
$x_{Bj}$ is not too close to 0 and 1) follows from the factorization
theorem (FT) considered by several authors (see e.g. refs.
\cite{ER,CSS,CSS1,Sterman}).

 The formulations of the FT given by different
authors differ each other but the idea of all the formulations is that

\begin{itemize}
\item $1_{FT}$) The amplitudes of lepton-parton interactions can be
calculated in the framework of perturbative QCD.

\item $2_{FT}$) The parton distribution functions (PDFs)
are essentially
nonperturbative objects, but they are universal in the sense that
they are the same in all hard-scattering processes.
\end{itemize}

The condition $1_{FT})$ follows from the property
that in amplitudes of the lepton-parton interactions entering into
diagrams dominating in DIS the off-shellness of quarks interacting
with the virtual photon, W or Z bosons is very large. This implies
that the (electromagnetic or weak) current operator ${\hat J}^{\mu}(x)$
(where $\mu=0,1,2,3$ and
$x$ is a point in Minkowski space) responsible for the transition
$nucleon \rightarrow hadrons$ in DIS can be considered in the
framework of perturbation theory.

 If $|N \rangle $ is the state of the initial nucleon then the DIS
cross-section is fully defined by the hadronic tensor
\begin{equation}
W^{\mu\nu}=\frac{1}{4\pi}\int\nolimits e^{\imath qx} \langle N|
{\hat J}^{\mu}(\frac{x}{2}){\hat J}^{\nu}(-\frac{x}{2})
|N \rangle d^4x
\label{1}
\end{equation}
The usual motivation of the FT is that since
the main contribution to the integral (\ref{1}) is given by small
distances (more precisely by the values of $x$ close to the light cone),
the possibility to consider the operator ${\hat J}^{\mu}(x)$
in the framework of perturbation theory is a consequence
of asymptotic freedom (which implies that $\alpha_s(Q^2)\rightarrow 0$
when $Q\rightarrow \infty$, where $\alpha_s$ is the QCD running coupling
constant). However a rigorous proof of the FT
encounters serious difficulties (see the discussion in ref.
\cite{CSS1}). As noted in ref. \cite{CSS}, "it is fair to say that
a rigorous treatment of factorization has yet to be provided".

 In any relativistic quantum theory the system under consideration is
described by some unitary representation $(a,l)\rightarrow {\hat U}(a,l)$
of the Poincare group. Here ${\hat U}(a,l)$ is the representation
operator corresponding to the element $(a,l)$ of the Poincare group,
$a$ is the four-vector describing the displacement of the origin
in spacetime translation of Minkowski space and $l\in SL(2,C)$.
We use ${\hat U}(a)$ to denote ${\hat U}(a,1)$ and ${\hat U}(l)$
to denote ${\hat U}(0,l)$. Let ${\hat P}=({\hat P}^0,{\hat {\bf
P}})$ be the four-momentum operator, where ${\hat P}^0$ is the
Hamiltonian and ${\hat {\bf P}}$ is the operator of ordinary
momentum. Then ${\hat U}(a)=exp(\imath {\hat P}_{\mu}a^{\mu})$
where a sum over repeated indices is assumed. Analogously it is
well-known that ${\hat U}(l)$ can be written in terms of the
operators ${\hat M}^{\mu\nu}$ (${\hat M}^{\mu\nu}=-{\hat M}^{\nu\mu}$)
which are the representation generators of the Lorentz group.

We shall always assume that the commutation relations for the
representation generators of the Poincare group are realized in the form
$$[{\hat P}^{\mu},{\hat P}^{\nu}]=0, \quad [{\hat M}^{\mu\nu},
{\hat P}^{\rho}]= -\imath({\eta}^{\mu\rho}{\hat P}^{\nu}-
{\eta}^{\nu\rho}{\hat P}^{\mu}),$$
\begin{equation}
[{\hat M}^{\mu\nu},{\hat M}^{\rho\sigma}]=-\imath ({\eta}^{\mu\rho}
{\hat M}^{\nu\sigma}+{\eta}^{\nu \sigma}{\hat M}^{\mu\rho}-{\eta}^
{\mu\sigma}{\hat M}^{\nu\rho}-{\eta}^{\nu\rho}{\hat M}^{\mu\sigma})
\label{2}
\end{equation}
where $\mu,\nu,\rho,\sigma=0,1,2,3$, the metric tensor in Minkowski
space has the nonzero components $\eta^{00}=-\eta^{11}=-\eta^{22}=
-\eta^{33}=1$, and we use the system of units with $\hbar=c=1$.

\begin{sloppypar}
 As explained in the well-known textbooks and monographs (see e.g.
refs. \cite{AB,BLOT}), matrix elements of field operators over
some states have correct transformation properties relative
transformations from the Poincare group only if transformations
of the operators are compatible with the transformations of the
states. This implies in particular that the hadronic tensor
$W^{\mu\nu}$ in Eq. (\ref{1}) will be the true tensor only if
the current operator is the relativistic vector operator
satisfying the conditions
\begin{equation}
{\hat U}(a)^{-1}{\hat J}^{\mu}(x){\hat U}(a)=
{\hat J}^{\mu}(x-a),
\label{3}
\end{equation}
\begin{equation}
{\hat U}(l)^{-1}{\hat J}^{\mu}(x){\hat U}(l)=L(l)^{\mu}_{\nu}
{\hat J}^{\nu}(L(l)^{-1}x)
\label{4}
\end{equation}
where $L(l)$ is the element of the Lorentz group corresponding to $l$.
The second expression implies that, as a consequence of Lorentz
invariance,
\begin{equation}
[{\hat M}^{\mu\nu},
{\hat J}^{\rho}(x)]= -\imath\{(x^{\mu}\partial^{\nu}-x^{\nu}\partial^{\mu})
{\hat J}^{\rho}(x)+\eta^{\mu\rho}{\hat J}^{\nu}(x)-\eta^{\nu\rho}
{\hat J}^{\mu}(x)\}
\label{5}
\end{equation}
\end{sloppypar}

 In addition, the current operator should be local in the sense that
$[{\hat J}^{\mu}(x/2),{\hat J}^{\nu}(-x/2)]=0$
if $x^2<0$.

\begin{sloppypar}
 Let us now consider the difference between relativistic invariance
and covariance although different authors understand these
notions differently. The latter often means that each stage of
calculations
involves only objects belonging to finite-dimensional representations
of the group SL(2,C) --- Dirac spinors, four-vectors, their
scalar products in Minkowski metric etc. The former means that all
operators corresponding to physical observables satisfy proper
commutation relations with the Poincare group representation operators
for the system under consideration. It is clear that in the general case
relativistic invariance is the necessary physical condition while
covariance is not.
\end{sloppypar}

 In QED the electrons, positrons and photons are the fundamental
particles, and the scattering space is the space of these almost
free particles. The operators ${\hat P}^{\mu},{\hat M}^{\mu\nu}$
act in the scattering space of the system under consideration,
and therefore, when the S-matrix is calculated in perturbation
theory, they can be replaced by the corresponding free operators
$P^{\mu},M^{\mu\nu}$. Then, as shown in the standard textbooks
(see e.g. ref. \cite{AB}), a covariant way of calculating the S-matrix
guarantees that it will be relativistically invariant.

However in QCD the scattering space by no means can be considered as a
space of almost free fundamental particles --- quarks and gluons. For
example, even if the scattering space consists of one particle (say the
nucleon), this particle is the bound state of quarks and gluons, and the
operators ${\hat P}^{\mu},{\hat M}^{\mu\nu}$ considerably differ from
$P^{\mu},M^{\mu\nu}$. It is well-known (even in the nonrelativistic
quantum mechanics) that in the presence of bound states the sets
(${\hat P}^{\mu},{\hat M}^{\mu\nu}$) and ($P^{\mu},M^{\mu\nu}$) cannot
be unitarily equivalent, and, since perturbation theory does not apply to
bound states, the former operators cannot be determined in the
framework of perturbation theory. In this case covariance does not
guarantee that the results will be relativistically invariant since
the operators in question should properly commute with
${\hat P}^{\mu},{\hat M}^{\mu\nu}$ and not with $P^{\mu},M^{\mu\nu}$.
The usual assumption in covariant calculations involving a
bound state is that any matrix element describing a transition of
this state can be (at least in principle) obtained by calculating
matrix elements involving only transitions of fundamental particles
into each other and taking the residue corresponding to the bound
state. However in view of confinement it is not clear how to
substantiate the scattering problem for free quarks and gluons.
For these reasons we will be interested in cases when the
representation operators in Eqs. (\ref{3}) and (\ref{4})
correspond to the full generators ${\hat P}^{\mu},{\hat M}^{\mu\nu}$.

 Now the question arises whether perturbative consideration of the
current operator in the Bjorken limit is compatible with the
correct transformation properties of the quantity $W^{\mu\nu}$.
Indeed, if the current operator is considered in perturbation
theory then the momentum and angular momentum operators in
Eqs. (\ref{3}) and (\ref{4}) are not ${\hat P}^{\mu},{\hat M}^{\mu\nu}$
but $P^{\mu},M^{\mu\nu}$. At the same time,
the Poincare transformations of the state $|N\rangle$ are described by
the operators ${\hat P}^{\mu},{\hat M}^{\mu\nu}$.

\begin{sloppypar}
 Another possible approach to DIS is as follows. Instead of the
FT we use the operator product expansion (OPE) \cite{Wil} according to
which

\begin{itemize}
\item $1_{OPE}$) The product of the operators in Eq. (\ref{1})
can be written in the form
\begin{equation}
{\hat J}(\frac{x}{2}){\hat J}(-\frac{x}{2})
= \sum_{i} C_i(x^2) x_{\mu_1}\cdots x_{\mu_n}
{\hat O}_i^{\mu_1\cdots \mu_n}
\label{6}
\end{equation}
where $C_i(x^2)$ are the $c$-number Wilson coefficients and the
operators ${\hat O}_i^{\mu_1\cdots \mu_n}$ are regular at $x=0$.

\item $2_{OPE}$) These operators depend only on field
operators and their covariant derivatives at the origin of Minkowski
space and have the same form as in perturbation theory.
\end{itemize}

For example, the basis for twist two operators contains in particular
\begin{equation}
{\hat O}_V^{\mu}={\cal N} \{{\hat {\bar \psi}}(0)\gamma^{\mu}{\hat
\psi}(0)\} \quad
{\hat O}_A^{\mu}={\cal N} \{{\hat {\bar \psi}}(0)\gamma^{\mu}
\gamma^5{\hat \psi}(0)\}
\label{7}
\end{equation}
where ${\cal N}$ stands for the normal product, ${\hat \psi}(x)$ is
the Heisenberg operator of the Dirac field and for simplicity we do not
write flavor operators and color and flavor indices.
\end{sloppypar}

However the OPE has been proved only in perturbation theory
of renormalized interactions \cite{Br}
while in view of Eqs. (\ref{3}) and (\ref{4}) we have to use the
OPE beyond perturbation theory. The problem of the validity of the
OPE beyond perturbation theory is rather difficult and so far concrete
results in this field have been obtained only in two-dimensional
models (see ref. \cite{Nov} and references therein).

Although there exists a vast literature devoted to the theory of
DIS, the restrictions imposed on the current operator by
Eqs. (\ref{3}) and (\ref{4}) have not been considered. In view of
the above discussion it seems reasonable to investigate whether
these restrictions can add something to the understanding of the
validity of the properties $1_{FT})$, $2_{FT})$, $1_{OPE})$ and
$2_{OPE})$.

The paper is organized as follows. In Sects. \ref{S2} and
\ref{S3} we discuss the general properties of the current operator in
quantum field theory and the properties derived in the framework of
canonical formalism. In addition to the results of many authors it is
shown in Sect. \ref{S4}, that the latter properties are not reliable since
in some cases they are incompatible with Lorentz invariance. The
current operator in DIS is considered in Sect. \ref{S5} and it is
argued that the nonperturbative part of this operator contributes
to DIS even in the Bjorken limit. In Sect. \ref{S6} we describe a
model which, in our opinion, is important for understanding the
problems considered in the paper. Finally Sect. \ref{S7} is
discussion.

\section{Problems with constructing the current operator}
\label{S2}

 Strictly speaking, the notion of current is not necessary if the theory
is complete. For example, in QED there exist unambiguous prescriptions
for calculating the elements of the S-matrix to any desired order of
perturbation theory and this is all we need. It is believed that this
notion is useful for describing the electromagnetic or weak properties
of strongly interacted systems. It is sufficient to know the matrix
elements $\langle \beta|{\hat J}^{\mu}(x)|\alpha \rangle$ of the
operator ${\hat J}^{\mu}(x)$ between the (generalized) eigenstates of the
operator ${\hat P}^{\mu}$ such that ${\hat P}^{\mu}|\alpha\rangle =
P_{\alpha}^{\mu}|\alpha\rangle$, ${\hat P}^{\mu}|\beta\rangle=
P_{\beta}^{\mu}|\beta\rangle$. It is usually assumed that as a
consequence of Eq. (\ref{3})
\begin{equation}
\langle \beta|{\hat J}^{\mu}(x)|\alpha \rangle=exp[\imath(P_{\beta}^{\nu}-
P_{\alpha}^{\nu})x_{\nu}] \langle \beta|{\hat J}^{\mu}|\alpha \rangle
\label{8}
\end{equation}
where formally ${\hat J}^{\mu}\equiv {\hat J}^{\mu}(0)$. Therefore
in the absence of a complete theory we can consider the less fundamental
problem of investigating the properties of the operator ${\hat J}^{\mu}$.
From the mathematical point of view this implies that we treat
${\hat J}^{\mu}(x)$ not as a four-dimensional operator distribution, but
as a usual operator function satisfying the condition
\begin{equation}
{\hat J}^{\mu}(x)=exp(\imath {\hat P}x){\hat J}^{\mu}
exp(-\imath {\hat P}x)
\label{9}
\end{equation}

\begin{sloppypar}
 The standpoint that the current operator should not be treated on the
same footing as the fundamental local fields is advocated by several
authors in their investigations on current algebra (see e.g.
ref. \cite{AFFR}). One of the arguments is that, for example, the
canonical current operator in QED is given by \cite{AB}
\begin{equation}
{\hat J}^{\mu}(x)={\cal N} \{{\hat {\bar \psi}}(x)\gamma^{\mu}{\hat
\psi}(x)\}=\frac{1}{2}[{\hat {\bar \psi}}(x),\gamma^{\mu}{\hat \psi}(x)]
\label{10}
\end{equation}
but this expression is not a
well-definition of a local operator. Indeed, as explained in several
well-known monographs (see e.g. refs. \cite{Str,BLOT}), the interacting
field operators can be treated only as operator valued distributions and
therefore the product of two local field operators at coinciding points
is not well defined. The problem of the correct definition of such
products is known as that of constructing composite operators (see e.g.
ref. \cite{Zim}).
So far this problem has been solved only in the framework
of perturbation theory for special models. When perturbation theory
does not apply the usual prescriptions are to separate the arguments
of the operators in question and to define the composite operator as
a limit of nonlocal operators when the separation goes to zero (see e.g.
ref. \cite{J} and references therein). Since we do not know how to
work with quantum field theory beyond perturbation theory, we do not
know what is the correct prescription.
\end{sloppypar}

 An additional difficulty with the current operator given by Eq.
(\ref{10}) is as follows. It is well-known that this operator is unitarily
equivalent to the free current operator (to the current operator
in interaction picture). At the same time, as noted above, in the
general case the operators $({\hat P}^{\mu},{\hat M}^{\mu\nu})$
are not unitarily equivalent to $(P^{\mu},M^{\mu\nu})$. Therefore
the problem arises whether the operator given by Eq. (\ref{10})
will satisfy proper commutation relations with the operators
$({\hat P}^{\mu},{\hat M}^{\mu\nu})$ in the presence of bound states.

 It is well-known (see e.g. ref. \cite{J}) that it is possible
to add to the current operator the term $\partial_{\nu}X^{\mu\nu}(x)$
where $X^{\mu\nu}(x)$ is some operator antisymmetric in $\mu$ and $\nu$.
However it is usually believed \cite{J} that the electromagnetic and
weak current operators of strongly interacted systems are given by the
canonical quark currents the form of which is similar to that in Eq.
(\ref{10}).

 If the operator ${\hat J}^{\mu}$ can be correctly defined then,
as follows from Eqs. (\ref{5}) and (\ref{9}),
\begin{equation}
[{\hat M}^{\mu\nu},
{\hat J}^{\rho}]= -\imath (\eta^{\mu\rho}{\hat J}^{\nu}-\eta^{\nu\rho}
{\hat J}^{\mu})
\label{11}
\end{equation}

\section{Canonical quantization and the forms of relativistic
dynamics}
\label{S3}

 In the standard formulation of quantum field theory the operators
${\hat P}_{\mu},{\hat M}_{\mu\nu}$ are given by
\begin{equation}
{\hat P}_{\mu}=\int\nolimits {\hat T}_{\mu}^{\nu}(x)d\sigma_{\nu}(x),\quad
{\hat M}_{\mu\nu}=\int\nolimits {\hat M}_{\mu\nu}^{\rho}(x)d\sigma_{\rho}(x)
\label{12}
\end{equation}
where ${\hat T}_{\mu}^{\nu}(x)$ and ${\hat M}_{\mu\nu}^{\rho}(x)$ are the
energy-momentum and angular momentum tensors and
$d\sigma_{\mu}(x)=\lambda_{\mu}\delta(\lambda x-\tau)d^4x$ is
the volume element of the space-like hypersurface defined by the time-like
vector $\lambda \quad (\lambda^2=1)$ and the evolution parameter $\tau$.
In turn, these tensors are fully defined by the classical Lagrangian
and the canonical commutation relations on the hypersurface
$\sigma_{\mu}(x)$. In this connection we note that in canonical
formalism the quantum fields are supposed to be distributions only
relative the three-dimensional variable characterizing the points of
$\sigma_{\mu}(x)$ while the dependence on the variable describing the
distance from $\sigma_{\mu}(x)$ is usual \cite{BLOT}.

 In spinor QED we define $V(x)=-L_{int}(x)=e{\hat J}^{\mu}(x)
{\hat A}_{\mu}(x)$, where $L_{int}(x)$ is the quantum interaction
Lagrangian, $e$ is the (bare) electron charge and ${\hat A}_{\mu}(x)$ is
the operator of the Maxwell field (let us note that if
${\hat J}^{\mu}(x)$ is treated as a composite operator then the product
of the operators entering into $V(x)$ should be correctly defined).

 At this stage it is not necessary to require that ${\hat J}^{\mu}(x)$ is
given by Eq. (\ref{10}), but the key assumption in the canonical
formulation of QED is that ${\hat J}^{\mu}(x)$ is
constructed only from ${\hat \psi}(x)$ (i.e. there is no dependence on
${\hat A}_{\mu}(x)$ and the derivatives of the fields ${\hat A}_{\mu}(x)$
and ${\hat \psi}(x)$). Then the canonical result derived in several
well-known textbooks and monographs (see e.g. ref. \cite{AB}) is
\begin{equation}
{\hat P}^{\mu}=P^{\mu}+\lambda^{\mu}\int\nolimits V(x)
\delta(\lambda x-\tau)d^4x
\label{13}
\end{equation}
\begin{equation}
{\hat M}^{\mu\nu}=M^{\mu\nu}+\int\nolimits V(x)
(x^{\nu}\lambda^{\mu}-x^{\mu}\lambda^{\nu})
\delta(\lambda x-\tau)d^4x
\label{14}
\end{equation}
 It is important to note that if $A^{\mu}(x)$, $J^{\mu}(x)$ and
$\psi(x)$ are the corresponding free operators then
${\hat A}^{\mu}(x)=A^{\mu}(x)$, ${\hat J}^{\mu}(x)=J^{\mu}(x)$ and
${\hat \psi}(x)=\psi(x)$ if $x\in \sigma_{\mu}(x)$.

  As pointed out by Dirac \cite{Dir}, any physical system can be
described in different forms of relativistic dynamics. By definition,
the description in the point form implies that the operators
${\hat U}(l)$ are the same as for noninteracting particles, i.e.
${\hat U}(l)=U(l)$ and ${\hat M}^{\mu\nu}=M^{\mu\nu}$, and thus
interaction terms can be present only in the four-momentum operators
${\hat P}$ (i.e. in the general case ${\hat P}^{\mu}\neq P^{\mu}$ for
all $\mu$). The description in the instant form implies that the
operators of ordinary momentum and angular momentum do not depend on
interactions, i.e. ${\hat {\bf P}}={\bf P}$, ${\hat {\bf M}}={\bf M}$
$({\hat {\bf M}}=({\hat M}^{23},{\hat M}^{31},{\hat M}^{12}))$, and
therefore interaction terms may be present only in ${\hat P}^0$ and the
generators of the Lorentz boosts ${\hat {\bf N}}=({\hat M}^{01},
{\hat M}^{02},{\hat M}^{03})$. In the front form with the marked $z$
axis we introduce the + and - components of the four-vectors as $x^+=
(x^0+x^z)/\sqrt{2}$, $x^-=(x^0-x^z)/\sqrt{2}$. Then we require that
the operators ${\hat P}^+,{\hat P}^j,{\hat M}^{12},{\hat M}^{+-},
{\hat M}^{+j}$ $(j=1,2)$ are the same as the corresponding free
operators, and therefore interaction terms may be present only in the
operators ${\hat M}^{-j}$ and ${\hat P}^-$.

 In quantum field theory the form of dynamics depends on the choice of
the hypersurface $\sigma_{\mu}(x)$. The representation generators of
the subgroup which leaves this hypersurface invariant are free since
transformations from this subgroup do not involve dynamics. Therefore
it is reasonable to expect that Eqs. (\ref{13}) and (\ref{14}) give the
most general form of the Poincare group representation generators in
quantum field theory if the fields are quantized on the hypersurface
$\sigma_{\mu}(x)$, but in the general case the relation between $V(x)$
and $L_{int}(x)$ is not so simple as in QED. The fact that the operators
$V(x)$ in Eqs. (\ref{13}) and (\ref{14}) are the same follows from
Eq. (\ref{2}).

 The most often considered case is $\tau =0$, $\lambda =(1,0,0,0)$. Then
$\delta(\lambda x-\tau)d^4x=d^3{\bf x}$ and the integration in Eqs.
(\ref{13}) and (\ref{14}) is taken over the hyperplane $x^0=0$. Therefore,
as follows from these expressions, ${\hat {\bf P}}={\bf P}$ and
${\hat {\bf M}}={\bf M}$. Hence such a choice of $\sigma_{\mu}(x)$
leads to the instant form \cite{Dir}.

  The front form can be formally obtained from Eqs. (\ref{13}) and
(\ref{14}) as follows. Consider the vector $\lambda$ with the components
\begin{equation}
\lambda^0=\frac{1}{(1-v^2)^{1/2}},\quad \lambda^j=0,\quad
\lambda^3=-\frac{v}{(1-v^2)^{1/2}}\quad (j=1,2)
\label{15}
\end{equation}
Then taking the limit $v\rightarrow 1$ in Eqs. (\ref{13}) and
(\ref{14}) we get
\begin{eqnarray}
&&{\hat P}^{\mu}=P^{\mu}+\omega^{\mu}\int\nolimits V(x)
\delta(x^+)d^4x,\nonumber\\
&&{\hat M}^{\mu\nu}=M^{\mu\nu}+\int\nolimits V(x)
(x^{\nu}\omega^{\mu}-x^{\mu}\omega^{\nu})
\delta(x^+)d^4x
\label{16}
\end{eqnarray}
where the vector $\omega$ has the components $\omega^-=1$,
$\omega^+=\omega^j=0$. It is obvious that the generators (\ref{16}) are
given in the front form and that's why Dirac \cite{Dir} related this form
to the choice of the light cone $x^+=0$.

 In ref. \cite{Dir} the point form was related to the hypersurface
$t^2-{\bf x}^2>0,\,t>0$, but as argued by Sokolov \cite{Sok},
the point form should be related to the hyperplane orthogonal to the
four-velocity of the
system under consideration. We shall not discuss this question in the
present paper.

 In the case of systems with a finite number of particles all the
forms are unitarily equivalent \cite{SoSh} and therefore the choice
of the form is only the matter of convenience but not the matter of
principle. However in quantum field theory it is not clear whether
there exist forms in which the Poincare group representation
operators are correctly defined \cite{Str,BLOT}.

 It is clear that when a form of dynamic is chosen, the expressions
for the representation operators and the wave functions become
noncovariant. Nevertheless, in the Feynman diagram technique each
term of the perturbative series for the S-matrix is covariant. In
QCD it is not clear whether covariance can be preserved in the
presence of bound states (see the discussion in Sect. \ref{S1})
but anyway, as noted above, covariance is not a necessary condition.

\begin{sloppypar}
\section{Incompatibility of canonical formalism with Lorentz invariance
for spinor fields}
\label{S4}
\end{sloppypar}

 It has been shown in a vast literature (see e.g. refs.
\cite{Schw,Bj,JL,AD,JP,J,MW}) that the canonical treatment of commutation
relations between the current operators can often lead to incorrect
results. The results of these references show that it is often
premature to trust assumptions which may seem physical or natural.
The results of the present section can be considered as an additional
argument in favor of this point of view. Namely, the purpose of this
section is to show that the relation ${\hat J}^{\mu}(x)=J^{\mu}(x)$
if $x\in \sigma_{\mu}(x)$, which is the key property of the current
operator for the spinor field in canonical formalism, is not correct
since it is incompatible with the correct commutation relations
between the operators ${\hat J}^{\mu}(x)$ and ${\hat M}^{\mu\nu}$
(i.e. with Lorentz invariance of the current operator).

 A possible objection against the derivation of Eqs. (\ref{13}) and
(\ref{14}) is that the product of local operators at
one and the same value of $x$ is not a well-defined object. For
example, if $x^0=0$ then following Schwinger \cite{Schw}, instead of
Eq. (\ref{10}), one can define $J^{\mu}({\bf x})$ as the limit of the
operator
\begin{equation}
J^{\mu}({\bf x})=\frac{1}{2}[{\bar \psi}({\bf x}+\frac{{\bf l}}{2}),
\gamma^{\mu}exp(\imath e
\int_{{\bf x}-\frac{{\bf l}}{2}}^{{\bf x}+\frac{{\bf l}}{2}}
{\bf A}({\bf x}')d{\bf x}')
 \psi ({\bf x}-\frac{{\bf l}}{2})]
\label{17}
\end{equation}
when ${\bf l}\rightarrow 0$, the limit should be taken only at the
final stage of calculations and in the general case the time
components of the arguments of ${\hat {\bar \psi}}$ and ${\hat \psi}$
also differ each other (the contour integral in this expression is
needed to conserve gauge invariance). Therefore there is a "hidden"
dependence of ${\hat J}^{\mu}(x)$ on ${\hat A}^{\mu}(x)$ and hence Eqs.
(\ref{13}) and (\ref{14}) are incorrect.

 However, any attempt to separate the arguments of the ${\hat \psi}$
operators in ${\hat J}^{\mu}(x)$ immediately results in breaking of
locality. In particular, at any ${\bf l}\neq 0$ in Eq. (\ref{17}) the
Lagrangian is nonlocal and the whole edifice of local quantum field
theory (including canonical
formalism) becomes useless. Meanwhile the only known way of constructing
the generators ${\hat P}^{\mu},{\hat M}^{\mu\nu}$ in local quantum
field theory is canonical formalism. For these reason we first consider
the results which formally follow from canonical formalism and
then show that they are inconsistent.

\begin{sloppypar}
 In addition to the properties discussed above, the current operator
should also satisfy the continuity equation $\partial
{\hat J}^{\mu}(x)/\partial x^{\mu}=0$. As follows from this equation and
Eq. (\ref{3}), $[{\hat  J}^{\mu}(x),{\hat P}_{\mu}]=0$.
The canonical formalism in the instant form implies that if $x^0=0$ then
${\hat  J}^{\mu}({\bf x})=J^{\mu}({\bf x})$. Since $J^{\mu}({\bf x})$
satisfies the condition $[J^{\mu}({\bf x}),P_{\mu}]=0$, it follows from
Eq. (\ref{13}) that if ${\hat P}^{\mu}=P^{\mu}+V^{\mu}$ then the
continuity equation is satisfied only if
\begin{equation}
[V^0,J^0({\bf x})]=0
\label{18}
\end{equation}
where
\begin{equation}
V^0=\int\nolimits V({\bf x})d^3{\bf x},\quad V({\bf x}) =-e
{\bf A}({\bf x}){\bf J}({\bf x})
\label{19}
\end{equation}
We take into account the fact that the canonical quantization on the
hypersurface $x^0=0$ implies that $A^0({\bf x})=0$.
\end{sloppypar}

 As follows from Eqs. (\ref{3}) and (\ref{5}), the commutation relation
between the operators ${\hat M}^{0i}$ $(i=1,2,3)$ and $J^0({\bf x})$
should have the form
\begin{equation}
[{\hat M}^{0i},J^0({\bf x})]=-x^i[{\hat P}^0,J^0({\bf x})]-
\imath J^i({\bf x})
\label{20}
\end{equation}
Since
\begin{equation}
[M^{0i},J^0({\bf x})]=-x^i[P^0,J^0({\bf x})]-\imath J^i({\bf x})
\label{21}
\end{equation}
it follows from Eqs. (\ref{14}), (\ref{18}) and (\ref{19}) that Eq.
(\ref{20}) is satisfied if
\begin{equation}
\int\nolimits y^i {\bf A}({\bf y})[{\bf J}({\bf y}),J^0({\bf x})]
d^3{\bf y}=0
\label{22}
\end{equation}
It is well-known that if the standard equal-time commutation relations
are used naively then the commutator in Eq. ({\ref{22}) vanishes and
therefore this equation is satisfied. However when ${\bf x}\rightarrow
{\bf y}$ this commutator involves the product of four Dirac fields at
${\bf x}={\bf y}$. The famous Schwinger result \cite{Schw} is that if
the current operators in question are given by Eq. (\ref{17}) then
\begin{equation}
[J^i({\bf y}),J^0({\bf x})]=C\frac{\partial}{\partial x^i}
\delta({\bf x}-{\bf y})
\label{23}
\end{equation}
where $C$ is some (infinite) constant. Therefore Eq. ({\ref{22}) is not
satisfied and the current operator ${\hat J}^{\mu}(x)$ constructed in
the framework of canonical formalism does not satisfy Lorentz invariance.

 At the same time, Eq. (\ref{23}) is compatible with Eqs. (\ref{18}) and
(\ref{19}) since $div({\bf A}({\bf x}))=0$. One can also expect that the
commutator $[{\hat M}^{0i},J^k({\bf x})]$ is compatible with Eq. (\ref{5}).
This follows from the fact \cite{GJ} that if Eq. (\ref{23}) is
satisfied then the commutator $[J^i({\bf x}),J^k({\bf y})]$ does not
contain derivatives of the delta function.

 While the arguments given in
ref. \cite{Schw} prove that the commutator in Eq. (\ref{23}) cannot vanish,
one might doubt whether the singularity of the commutator is indeed given
by the right hand side of this expression. Of course, at present any
method of calculating such a commutator is model dependent, but
the incompatibility of canonical formalism with Lorentz invariance (see
Eq. (\ref{20}))
follows in fact only from algebraic considerations. Indeed, Eqs.
(\ref{18}), (\ref{20}) and (\ref{21}) imply that if ${\hat M}^{\mu\nu}=
M^{\mu\nu}+V^{\mu\nu}$ then
\begin{equation}
[V^{0i},J^0({\bf x})]=0
\label{24}
\end{equation}

 Since $V^{0i}$ in the instant form is a nontrivial interaction
dependent operator, there is no reason to expect that it commutes with
the free operator $J^0({\bf x})$. Moreover for the analogous reason
Eq. (\ref{18}) will not be satisfied in the general case.

 To better understand the situation in spinor QED it is useful to
consider scalar QED \cite{J2}. The formulation of this theory can
be found, for example, in ref. \cite{IZ}. In contrast with spinor
QED, the Schwinger term in scalar QED emerges canonically \cite{Schw,J}.
We use $\varphi({\bf x})$ to denote the operator of the scalar complex
field at $x^0=0$. The canonical calculation yields
\begin{eqnarray}
&&{\hat J}^0({\bf x})=J^0({\bf x})=\imath[\varphi^*({\bf x}) \pi^*({\bf x})-
\pi({\bf x})\varphi({\bf x})], \nonumber\\
&& {\hat J}^i({\bf x})=J^i({\bf x})-
2eA^i({\bf x})\varphi^*({\bf x}) \varphi({\bf x}),\nonumber\\
&& J^i({\bf x})=\imath [\varphi^*({\bf x})\cdot\partial^i
\varphi({\bf x})-\partial^i \varphi^*({\bf x})\cdot
\varphi({\bf x})]
\label{25}
\end{eqnarray}
where $\pi({\bf x})$ and $\pi^*({\bf x})$ are the operators canonically
conjugated with $\varphi({\bf x})$ and $\varphi^*({\bf x})$ respectively.
In contrast with Eq. (\ref{19}), the operator $V({\bf x})$ in scalar
QED is given by
\begin{equation}
V({\bf x}) =-e {\bf A}({\bf x}){\bf J}({\bf x})
+e^2{\bf A}({\bf x})^2\varphi^*({\bf x})\varphi({\bf x})
\label{26}
\end{equation}
However the last term in this expression does not contribute to the
commutator (\ref{20}). It is easy to demonstrate that as pointed out
in ref. \cite{J2}, the commutation relations (\ref{5}) in scalar QED
are satisfied in the framework of the canonical formalism.
Therefore the naive treatment of the product of local operators at
coinciding points in this theory is not in conflict with the
canonical commutation relations.
The key difference between spinor QED and scalar QED is that, in contrast
with spinor QED, the spatial component of the canonical current operator
is not free if $x^0=0$ (see Eq. (\ref{25})). Just for this
reason the commutator $[{\hat M}^{0i},J^0({\bf x})]$ in scalar QED
agrees with Eq. ({\ref{5}) since the Schwinger term in this commutator
gives the interaction term in ${\hat J}^i({\bf x})$.

 Now let us return to spinor QED. As noted above, the canonical
formalism cannot be used if the current operator is considered as a
limit of the expression similar to that in Eq. (\ref{17}). In
addition, the problem exists what is the correct definition of
$V({\bf x})$ as a composite operator. One might expect that the
correct definition of $J^{\mu}({\bf x})$ and $V({\bf x})$ will result
in appearance of some additional terms in $V({\bf x})$ (and hence in $V^0$
and $V^{0i}$). However it is unlikely that this is the main reason of
the violation of Lorentz invariance. Indeed, as noted above, for only
algebraic reasons it is unlikely that both conditions (\ref{18}) and
(\ref{24}) can be simultaneously satisfied. Therefore, taking into
account the situation in scalar QED, {\it it is natural to think that
the main reason of the failure of canonical formalism is that either the
limit of ${\hat J}^{\mu}(x^0,{\bf x})$ when $x^0\rightarrow 0$ does not
exist or this limit is not equal to $J^{\mu}({\bf x})$} (i.e. the
relation ${\hat J}^{\mu}({\bf x})=J^{\mu}({\bf x})$ is incorrect).

 The fact that the relation ${\hat J}^{\mu}({\bf x})=J^{\mu}({\bf x})$
cannot be correct follows from simpler considerations. Indeed, assume
first that this relation is valid. Then we can use canonical formalism
in the framework of which the generator of the gauge transformations is
$div {\bf E}({\bf y}) - J^0({\bf y})$, and if ${\bf J}({\bf x})$ is gauge
invariant then $[div {\bf E}({\bf y}) - J^0({\bf y}),
{\bf J}({\bf x})]=0$. The commutator $[J^0({\bf y}),{\bf J}({\bf x})]$
cannot be equal to zero \cite{Schw} and therefore ${\bf J}({\bf x})$
does not commute with $div {\bf E}({\bf y})$ while the free operator
${\bf J}({\bf x})$ commutes with $div {\bf E}({\bf y})$.
The relation ${\hat J}^{\mu}({\bf x})=J^{\mu}({\bf x})$ also does not
take place in explicitly solvable two-dimensional models \cite{BLOT}.
In addition, once we assume that the field operators on
the hypersurface $\sigma_{\mu}(x)$ are free we immediately are in
conflict with the Haag theorem \cite{Haag,Str,BLOT}. However for our
analysis of the current operator in DIS in Sect. \ref{S5} it is
important that ${\hat J}^{\mu}({\bf x})\neq J^{\mu}({\bf x})$ as a
consequence of Lorentz invariance.

\begin{sloppypar}
By analogy with ref. \cite{Schw} it is easy to show that if $x^+=0$
then the canonical current operator in the front form $J^+(x^-,
{\bf x}_{\bot})$ (we use the subscript $\bot$ to denote the projection
of the three-dimensional vector onto the plane 12) cannot commute with
all the operators $J^i(x^-,{\bf x}_{\bot})$ $(i=-,1,2)$. As easily
follows from the continuity equation and Lorentz invariance (\ref{5}),
the canonical operator $J^+(x^-,{\bf x}_{\bot})$ should satisfy the
relations
\begin{equation}
[V^-,J^+(x^-,{\bf x}_{\bot})]=[V^{-j},J^+(x^-,{\bf x}_{\bot})]=0
\quad (j=1,2)
\label{27}
\end{equation}
By analogy with the above consideration it is natural to think that these
relations cannot be simultaneously satisfied and therefore
{\it either the limit of ${\hat J}^{\mu}(x^+,x^-,{\bf x}_{\bot})$ when
$x^+\rightarrow 0$ does not exist or this limit is not equal to
$J^{\mu}(x^-,{\bf x}_{\bot})$}. Therefore the canonical light cone
quantization does not render a Lorentz invariant current operator for
spinor fields.
\end{sloppypar}

 Let us also note that if the theory should be invariant under the space
reflection or time reversal, the corresponding representation operators
in the front form ${\hat U}_P$ and ${\hat U}_T$ are necessarily
interaction dependent (this is clear, for example, from the relations
${\hat U}_PP^+{\hat U}_P^{-1}$ = ${\hat U}_TP^+{\hat U}_T^{-1}$ =
${\hat P}^-$). In terms of the operator ${\hat J}^{\mu}$ one can say
that this operator should satisfy the conditions
\begin{equation}
{\hat U}_P({\hat J}^0,{\hat {\bf J}}){\hat U}_P^{-1}=
{\hat U}_T({\hat J}^0,{\hat {\bf J}}){\hat U}_T^{-1}=
({\hat J}^0,-{\hat {\bf J}})
\label{28}
\end{equation}
Therefore it is not clear whether these conditions are compatible with
the relation ${\hat J}^{\mu}=J^{\mu}$. However this difficulty is
a consequence of the difficulty with Eq. (\ref{11}) since, as noted by
Coester \cite{Coes}, the interaction dependence of the operators
${\hat U}_P$ and ${\hat U}_T$ in the front form does not mean that there
are discrete dynamical symmetries in addition to the rotations about
transverse axes.
Indeed, the discrete transformation $P_2$ such that
$P_2\, x:= \{x^0,x_1,-x_2,x_3\}$ leaves the light front $x^+=0$ invariant.
The full space reflection $P$ is the product of $P_2$ and a rotation about
the 2-axis by $\pi$. Thus it is not an independent dynamical transformation
in addition to the rotations about transverse axes.
Similarly the transformation $TP$ leaves $x^+=0$ invariant and
$T=(TP)P_2R_2(\pi)$.

 In terms of the operator ${\hat J}^{\mu}$ the results of this section
are obvious. Indeed, since at $x=0$ the Heisenberg and Schroedinger
pictures coincide then in view of Eq. (\ref{10}) one might think that
the operator ${\hat J}^{\mu}$ is free, i.e. ${\hat J}^{\mu}=J^{\mu}$.
However there is no reason for the interaction terms in
${\hat M}^{\mu\nu}$
to commute with all the operators $J^{\mu}$ (see Eq. (\ref{11})).
Therefore the results of this section show that the algebraic reasons
based on Eq. (\ref{11}) are more solid than the reasons based on
formal manipulations with local operators and in the instant and front
forms ${\hat J}^{\mu}\neq J^{\mu}$ (moreover, ${\hat J}^{\mu}(x)
\neq J^{\mu}(x)$ if $x\in \sigma_{\mu}(x)$). In particular, if some
interaction operators are present in ${\hat M}^{\mu\nu}$, one has to
expect that they are also present in some of the operators
${\hat J}^{\mu}$.

 Let us note that although the model
considered in this section is spinor QED, the above results are not
very important for QED itself and other standard theories where
the S-matrix can be calculated in
perturbation theory. The matter is that, as pointed out in
Sect. \ref{S1}, in this case it is sufficient to consider only
commutators involving $P^{\mu}$ and $M^{\mu\nu}$. For example, the
problem important for calculating the S-matrix elements in such
theories is that of constructing the covariant $T$-product
$T^*({\hat J}^{\mu}(x){\hat J}^{\nu}(y))$ (see e.g.
refs. \cite{Bj,JL,AB,IZ}). It has been shown that in the presence of
Schwinger terms the standard $T$-product
$T({\hat J}^{\mu}(x){\hat J}^{\nu}(y))$ is not covariant
but it is possible to add to $T({\hat J}^{\mu}(x){\hat J}^{\nu}(y))$
a contact term (which is not equal to zero only if $x^0=y^0$) such
that the resulting $T^*$-product will be covariant. In standard
theories covariance is equivalent to Lorentz invariance since
covariance is satisfied when the operators in question properly
commute with the $M^{\mu\nu}$. However in QCD
Lorentz invariance is satisfied only if the operators in question
properly commute with the full operators ${\hat M}^{\mu\nu}$.
Therefore in this case Lorentz invariance and covariance differ each
other and the former is of course more important.

 In the next section we argue that the algebraic consideration
based on Eq. (\ref{11}) is important for investigating the properties
of the current operator for strongly interacting particles.

\section{Current operator in DIS}
\label{S5}

 The DIS cross-section is fully defined by the hadronic tensor
given by Eq. (\ref{1}). The nucleon state $|N\rangle$ is
the eigenstate of the operator ${\hat P}$ with the eigenvalue
$P'$ and the eigenstate of the spin operators ${\hat {\bf S}}^2$ and
${\hat S}^z$ which are constructed from ${\hat M}^{\mu\nu}$. In
particular, ${\hat P}^2|N\rangle =m^2 |N\rangle$ where $m$ is the
nucleon mass.

 The structure of the four-momentum operator ${\hat P}$ in QCD is rather
complicated (see e.g. ref. \cite{CCJ}) but anyway some of the
components of ${\hat P}$ necessarily contain a part which describes the
interaction of quarks
and gluons at large distances where $\alpha_s$ is large and
perturbation theory does not apply. In view of the last relation, this
part is responsible for binding of quarks and gluons in the nucleon.
We will call this part the nonperturbative one.

 Suppose that the Hamiltonian ${\hat P}^0$ contains the nonperturbative
part and consider the relation
$[{\hat M}^{0i},{\hat P}^k]=-\imath \delta_{ik}{\hat P}^0$ $(i,k=1,2,3)$
which follows from Eq. (\ref{2}).
Then it is obvious that if ${\hat P}^k=P^k$ then all the operators
${\hat M}^{0i}$ inevitably depend on the nonperturbative part and
{\it vice versa}, if ${\hat M}^{0i}=M^{0i}$ then all the operators
${\hat P}^k$ inevitably depend on this part.
Therefore in the instant form all the operators ${\hat M}^{0i}$
inevitably depend on the nonperturbative part and in the point form all
the operators ${\hat P}^k$
inevitably depend on this part. In the front form the fact that all the
operators ${\hat M}^{-j}$  inevitably depend on the
nonperturbative part follows from the relation
$[{\hat M}^{-j},{\hat P}^l]=-\imath \delta_{jl}{\hat P}^-$ $(j,l=1,2)$
which also is a consequence of Eq. (\ref{2}).
Of course, the physical results should not depend on the choice of the
form of dynamics and in the general case all ten generators can
depend on the nonperturbative part.

 In turn, Eq. (\ref{5}) and the consideration in Sect. \ref{S4},
give grounds to think that some of
the operators ${\hat J}^{\mu}({\bf x})$ in the instant form and
${\hat J}^{\mu}(x^-,{\bf x}_{\bot})$ in the front one
inevitably depend on the nonperturbative part. If it is possible to
define ${\hat J}^{\mu}$ in the
point form then as follows from Eq. (\ref{11}), the relation
${\hat J}^{\mu}=J^{\mu}$ does not contradict Lorentz invariance
but, as follows from Eq. (\ref{9}), the operator ${\hat J}^{\mu}(x)$ in
that form inevitably depend on the nonperturbative part.
As noted in Sect. \ref{S1}, the fact that the same operators
$({\hat P}^{\mu},{\hat M}^{\mu\nu})$
describe the transformations of both the operator ${\hat J}^{\mu}(x)$
and the state $|N\rangle$ guaranties that $W^{\mu\nu}$ has the
correct transformation properties.

 We see that the relation between the current operator and the state of
the initial nucleon is highly nontrivial. Meanwhile, in the present theory
they are considered separately. As noted in Sect. \ref{S1},
the possibility of the separate consideration
follows from the FT.

 Let us now discuss the following question. If the current
operator depends on the nonperturbative part then this operator depends
on the integrals from the quark
and gluon field operators over the region of large distances where
$\alpha_s$ is large. Is this property
compatible with locality? In the framework of canonical formalism
compatibility is obvious but, as shown in the preceding section,
the results based on canonical formalism are not reliable.
Therefore it is not clear whether in QCD it is possible to construct
local electromagnetic and weak current operators beyond perturbation
theory.  We will now consider whether such a possibility can be
substantiated in the framework of the OPE.

\begin{sloppypar}
 As noted above, there are grounds to think that the operator
${\hat J}^{\mu}(x)$ necessarily depends on the nonperturbative part while
Eq. (\ref{6}) has been proved only in perturbation
theory. Therefore if we use Eq. (\ref{6}) in DIS we have to
assume that either nonperturbative effects are not important to some
orders in $1/Q$ and then we can use Eq. (\ref{6}) only
to these orders (see e.g. ref. \cite{Jaffe}) or it is possible to use
Eq. (\ref{6}) beyond perturbation theory. The question also arises whether
Eq. (\ref{6}) is valid in all admissible forms of dynamics (as it should
be if it is an exact operator equality) or only in some forms.
\end{sloppypar}

 In the point form all the components of ${\hat P}$ depend on the
nonperturbative part
and therefore, in view of Eqs. (\ref{3}) or (\ref{9}), it is not clear
why there is no nonperturbative part in the $x$ dependence of the right
hand side of Eq. (\ref{6}), or if (for some reasons) it is possible to
include the nonperturbative part only into the operators ${\hat O}_i$
then why they have the same form as in perturbation theory.

 One might think that in the front form the $C_i(x^2)$ will be the
same as in perturbation theory, at least in the case when the process
is considered in the infinite momentum frame (IMF)
where the initial nucleon has a
large positive momentum along the $z$ axis. Then
the value of $q^-$ in DIS is very
large and therefore only a small vicinity of the light cone $x^+=0$
contributes to the integral (\ref{1}). Therefore we indeed have the
description in the front form where
the only dynamical component of ${\hat P}$ is ${\hat P}^-$.
Since the eigenvalues of ${\hat P}^-$ in the IMF are small,
the dependence of ${\hat P}^-$ on the nonperturbative part is of no
importance. Since there exist final hadrons (which are the bound states
of quarks and gluons) moving in the negative direction of the $z$ axis
in the IMF, the problem arises whether the final state interaction
(FSI) can be neglected. In addition, it is necessary to take into
account that the
operator ${\hat J}^{\mu}(x^-,{\bf x}_{\bot})$ in the front form
depends on the nonperturbative part. Nevertheless we
assume that Eq. (\ref{6}) in the front form is valid (see also the
consideration in the next section).

 If we assume as usual that there is no problem with the convergence
of the OPE series then experiment makes it possible to measure
each matrix element
$\langle N|{\hat O}_i^{\mu_1\cdots \mu_n}|N\rangle$
which should have correct transformation properties (if only the
series as a whole has the correct properties then the decomposition
(\ref{6}) is of no practical importance).
Let us consider, for example, the matrix element
$\langle N|{\hat O}_V^{\mu}|N\rangle$. By analogy with the
consideration in Sect. \ref{S1} we conclude that this matrix
element transforms as a four-vector
if the Lorentz transformations of ${\hat O}_V^{\mu}$ are described by the
operators ${\hat M}^{\mu\nu}$ describing the transformations of
$|N\rangle$, or in other words, by analogy with Eq. (\ref{11}),
\begin{equation}
[{\hat M}^{\mu\nu},{\hat O}_V^{\rho}]=-\imath (\eta^{\mu\rho}{\hat
O}_V^{\nu}-\eta^{\nu\rho} {\hat O}_V^{\mu})
\label{29}
\end{equation}
It is also clear that Eq. (\ref{29}) follows from Eqs.
(\ref{2}-\ref{7}).
Since the ${\hat M}^{-j}$ in the front form depend on the
nonperturbative part, the results of
Sect. \ref{S4} give grounds to think that at least
some components ${\hat O}_V^{\mu}$, and analogously some components
${\hat O}_i^{\mu_1\cdot \mu_n}$, also depend on the
nonperturbative part. Since Eq. (\ref{29}) does not contain any $x$ or
$q$ dependence, this conclusion has nothing to do with asymptotic
freedom and is valid even in leading order in $1/Q$. Therefore
the problem arises whether this fact is compatible with the
FT \cite{ER,CSS,CSS1,Sterman}).

 If the operators ${\hat O}_i^{\mu_1...\mu_n}$ depend on the
nonperturbative part, then by analogy
with the above considerations we conclude that the operators in
Eq. (\ref{7}) are ill-defined and the correct expressions for them
involve integrals from the field operators over large distances where
the QCD coupling constant is large. Therefore it is not clear whether
the operators ${\hat O}_i^{\mu_1...\mu_n}$ are local and whether the
Taylor expansion at $x=0$ is correct, but even if it is, the
expressions for ${\hat O}_i^{\mu_1...\mu_n}$ will depend on higher
twist operators which contribute even in leading order in $1/Q$.

 If (for some reasons) Eq. (\ref{6}) is valid and
no form of the operators ${\hat O}_i^{\mu_1...\mu_n}$ is prescribed
then all the standard results concerning the $Q^2$ evolution of
the structure functions remain. Indeed the only information
about the operators ${\hat O}_i^{\mu_1...\mu_n}$ we need is their tensor
structure since we should correctly parametrize the matrix elements
$\langle N|{\hat O}_i^{\mu_1\cdots \mu_n}|N\rangle$.
However the form of the operators ${\hat O}_i^{\mu_1...\mu_n}$
is important for the derivation of sum rules in DIS. We will discuss
this question below.

\section{A model}
\label{S6}

 The above discussion shows that the compatibility of Poincare
invariance of the current operator with the FT
and OPE is the difficult problem of the present theory. For this
reason it is interesting to consider models in which
the well-defined current operator and the representation generators
of the Poincare group can be explicitly constructed
and one can explicitly verify that they satisfy the commutation
relations (\ref{2}-\ref{5}). In this section we consider a model
the detailed description of which will be given elsewhere \cite{LPS}.

 Consider the wave function of a system of $n$ particles with the
four-momenta $p_i$, and masses $m_i$ ($i=1,2,...n$). Since $p_i^2=m_i^2$,
only three components of $p_i$ are independent and we choose
${\bf p}_{i\bot},p_i^+$ as these components.
Instead of the individual variables
$({\bf p}_{1\bot},p_1^+,...{\bf p}_{n\bot},p_n^+)$ we introduce the
$\bot$ and $+$ components of the total momentum
\begin{equation}
{\bf P}_{\bot}={\bf p}_{1\bot}+...+{\bf p}_{n\bot},\quad
P^+=p_1^+...+p_n^+
\label{30}
\end{equation}
and the internal (Sudakov) variables
\begin{equation}
\xi_i=\frac{p_i^+}{P^+},\quad {\bf k}_{i\bot}=
{\bf p}_{i\bot}-\xi_i{\bf P}_{\bot}
\label{31}
\end{equation}

 Let  us define the "internal" space ${\cal H}_{int}$ as the space of
functions $\chi({\bf k}_{1\bot},\xi_1,...{\bf k}_{n\bot},\xi_n)$
such that
\begin{equation}
||\chi||^2=\int\nolimits
|\chi({\bf k}_{1\bot},\xi_1,...{\bf k}_{n\bot},\xi_n)|^2
d\rho (int)\quad < \infty
\label{32}
\end{equation}
where $d\rho(int)$ is the volume element in the internal momentum
space and the spin variables are dropped for simplicity
(the explicit expression for $d\rho(int)$ is of no importance for
us). Then the full Hilbert space ${\cal H}$ can be realized
as the space of functions $\phi ({\bf P}_{\bot},P^+)$ with the range
in ${\cal H}_{int}$ and such that
\begin{equation}
\int\nolimits ||\phi({\bf P}_{\bot},P^+)||^2
\frac{d^2{\bf P}_{\bot}dP^+}{2(2\pi)^3P^+}\quad < \infty
\label{33}
\end{equation}

 Suppose that the particles interact with each other, and their
interactions are described in the front form. Then the $\bot$ and
$+$ components of the four-momentum operator are the operators of
multiplication by the corresponding variable, i.e.
${\hat {\bf P}}_{\bot}={\bf P}_{\bot}$, ${\hat P}^+=P^+$. As shown
by several authors (see e.g. refs. \cite{BT}), there exists the
representation where the remaining seven generators have the form
\begin{eqnarray}
&&{\hat P}^-=\frac{{\hat M}^2+{\bf P}_{\bot}^2}{2P^+}, \quad
M^{+-}=\imath P^+\frac{\partial}{\partial P^+},\nonumber\\
&&M^{+j}=-\imath P^+\frac{\partial}{\partial P^j},\quad
M^{xy}=-\imath \epsilon_{jl}P_j\frac{\partial}{\partial P_l}+S^z,\nonumber\\
&& {\hat M}^{-j}=-\imath(P^j\frac{\partial}{\partial P^+}+
{\hat P}^-\frac{\partial}{\partial P^j})-
\frac{\epsilon_{jl}}{P^+}({\hat M}{\hat S}^l+P^lS^z)
\label{34}
\end{eqnarray}
Here $j=(x,y)$, $\epsilon_{xx}=\epsilon_{yy}=0$,
$\epsilon_{xy}=-\epsilon_{yx}=1$,
${\hat M}$ is the mass operator and ${\hat {\bf S}}$ is the spin operator.
The latter act only through the variables of the space ${\cal H}_{int}$.

Any interacting system should satisfy cluster separability which has
been discussed by several
authors (see e.g. refs. \cite{sok1,sok,CP,M,lev1}). We will use the
algebraic version of cluster separability:
for any partition of the system under consideration into the
subsystems $(\alpha_1,...\alpha_k)$ such that all interactions
between these subsystems are turned off, the representation of the
Poincare group for the system as a whole becomes the tensor product
of the representations for the subsystems. This implies that the
representation generators for the system as a whole become sums of the
corresponding generators for the subsystems.

 As shown in ref. \cite{lev2}, a consequence of the cluster separability
property is that the spin operator is not equal to the free one;
only the $z$ component of ${\hat {\bf S}}$ is free while
${\hat {\bf S}}_{\bot}\neq {\bf S}_{\bot}$. However in models where
$n$ is finite
\begin{equation}
{\hat {\bf S}}=A {\bf S}A^{-1}
\label{35}
\end{equation}
where the unitary operator $A$ is the front form analog of the Sokolov
packing operators \cite{sok}.

 The explicit expression for $A$ is rather complicated and
essentially model dependent \cite{lev2}. In addition, the operator
satisfying Eq. (\ref{35}) is not unique since it is defined up to
unitary operators commuting with ${\bf S}$.

 The final step specifying our model is as follows. Since the
expressions (\ref{33}-{35}) do
not explicitly depend on $n$, we assume that field theory models
also can be described in such a way. In this case, taking into
account the experience with the finite values of $n$, we have
grounds to believe that the operator $A$ is highly nontrivial
and can be determined only beyond perturbation theory.

If $\chi$ is the internal nucleon wave function and $P'$ is its
total four-momentum, then its wave function in the representation
(\ref{34}) has the form
\begin{equation}
|N\rangle =2(2\pi)^3P^{'+}\delta^{(2)}({\bf P}_{\bot}-
{\bf P}_{\bot}')\delta(P^+-P^{'+})\chi
\label{36}
\end{equation}

 As follows from Eq. (\ref{9}), Eq. (\ref{1}) can be written in the
form
\begin{equation}
W^{\mu\nu}=\frac{1}{4\pi}\sum_{X} (2\pi)^4
\delta^{(4)}(P'+q-P_X) \langle N|{\hat J}^{\mu}|X\rangle
\langle X|{\hat J}^{\nu}|N\rangle
\label{37}
\end{equation}
where the sum is taken over all possible final states $X$ with the
four-momenta $P_X$.

We will consider the matrix elements of the current operator in the
reference frames where $P_z'$ is positive and very large. Then
$P^{'-}$ is very small and therefore the value of $P'$ is
approximately the same as in the case of the free momentum operator.
In the Bjorken limit it is also possible to find such frames
where $|P_X|\gg m$. We suppose as usual
that the FSI of the struck quark with the remnants of the target is
a higher twist effect, i.e. the effect which is suppressed as
$(m/Q)^{2n}\, (n=1,2...)$. Then there exist reference frames where
the values of $P_X$ also are approximately the same as for the free
four-momentum operator.

 As follows from Eq. (\ref{11})
\begin{equation}
[{\hat M}^{-j},{\hat J}^+]=-\imath {\hat J}^j,\quad
[{\hat M}^{-j},{\hat J}^l]=-\imath \delta_{jl}{\hat J}^-
\label{38}
\end{equation}
These expressions make it possible to obtain all the components of
${\hat J}^{\mu}$ if ${\hat J}^+$ is known.

 Suppose that in the Bjorken limit the action of ${\hat J}^+$ in the
reference frames under consideration can be replaced by the action
of the free operator $J^+$ and consider first the case when
$A=1$. The action of ${\hat M}$ on
$|X\rangle$ can be replaced by the action of $M$ and
therefore the interaction dependence of the operators ${\hat M}^{-j}$
manifests itself only when ${\hat M}$ acts on $|N\rangle$.
However since $m/P^{'+}$ is small, this dependence can be neglected.
We conclude that, as a consequence of the first expression in Eq.
(\ref{38}), the matrix elements of the operator ${\hat J}^j$ in the
reference frames under consideration can be calculated with
$J^j$. Analogously, as a consequence of the second expression,
the matrix elements of ${\hat J}^-$ can be calculated with $J^-$.

 We see that if $A=1$
then in the reference frames described above the matrix elements of
${\hat J}^{\mu}$  can be calculated with the operator $J^{\mu}$
and the interaction dependence of the four-momentum operator can be
neglected too. Therefore the tensor (\ref{1}) can be calculated with
the operator $J^{\mu}(x)$.

 However, as explained above, the
interaction dependence of the operators ${\hat M}^{-j}$ comes not only from
the interaction dependence of the mass operator but also from
the interaction dependence of the operator ${\bf S}$. In this case
the operators ${\hat M}^{-j}$ are unitarily equivalent to the operators
with $A=1$. Then the calculation of matrix elements in the
reference frames where ${\bf P}'$ and ${\bf P}_X$ are directed along the
$z$ axis by using $J^{\mu}(x)$ does not contradict Poincare
invariance, but it is not clear where the current operator as a whole
is local \cite{LPS}. At the same time, since $A$ becomes unity when
all interactions in the system under consideration are turned off,
it is obvious that the matrix
elements of the operator ${\hat J}^{\mu}(x)$ in the reference frames under
consideration can be calculated with the operator
\begin{equation}
{\hat J}^{\mu}(x)=AJ^{\mu}(x)A^{-1}
\label{39}
\end{equation}
which is local since $J^{\mu}(x)$ is local.

\begin{sloppypar}
 As noted in Sect. \ref{S1}, the operators considered in perturbation
theory properly commute
with the free representation generators of the Poincare group.
Therefore the obvious generalization of Eq. (\ref{39}) is
\begin{equation}
{\hat J}^{\mu}(x)=AJ_{pert}^{\mu}(x)A^{-1}
\label{40}
\end{equation}
where the operator $J_{pert}^{\mu}$ can be calculated in perturbative
QCD.
\end{sloppypar}

 Let us stress that Eqs. (\ref{39}) and (\ref{40}) are considered
not as exact operator equalities but only as the relations which can
be used if the matrix elements in Eq. (\ref{37}) are calculated in
some reference frames (for example, these relations cannot be used
in the reference frames where $P_z'$ is negative and $|P_z'|$ is
large).

 As already mentioned, the OPE has been proved only in perturbation
theory but several authors tried to prove the OPE beyond that
theory in 1+1 dimensional models \cite{Nov}. Let us note in
this connection that in the $(t,z)$
space the operator ${\hat {\bf S}}$ is absent (moreover, the
Lorentz group is one-dimensional and the only generator of this
group is $M^{+-}$) and therefore we can choose $A=1$. Hence the
validity of the OPE beyond perturbation theory in 1+1
dimensions does not necessarily imply that the same takes place
in 3+1 dimensions.

 As follows from Eq. (\ref{40}),
\begin{equation}
{\hat J}^{\mu}(\frac{x}{2}){\hat J}^{\nu}(-\frac{x}{2})=
AJ_{pert}^{\mu}(\frac{x}{2})J_{pert}^{\nu}(-\frac{x}{2})A^{-1}
\label{41}
\end{equation}
Now we apply the OPE to $J_{pert}^{\mu}(x/2)
J_{pert}^{\nu}(-x/2)$:
\begin{equation}
J_{pert}^{\mu}(\frac{x}{2})J_{pert}^{\nu}(-\frac{x}{2})
= \sum_{i} C_i(x^2) x_{\mu_1}\cdots x_{\mu_n}
{\tilde O}_i^{\mu_1\cdots \mu_n}
\label{42}
\end{equation}
where the operators ${\tilde O}_i^{\mu_1\cdots \mu_n}$ depend only on field
operators and their covariant derivatives at the origin of Minkowski
space. As follows from Eqs. (\ref{1}), (\ref{41}) and (\ref{42}),
the $Q^2$ evolution of the structure functions is defined by the
Wilson coefficients as well as in the standard theory.

 Instead of $\chi$ we now introduce a new
internal wave function ${\tilde \chi}$ such that
$\chi=A{\tilde \chi}$. Let us also introduce the functions
\begin{eqnarray}
&&\rho_i(\xi_i)=\sum_{\sigma_i}\int\nolimits
|\chi(\xi_i,{\bf k}_{i\bot},\sigma_i,int_i)|^2
\frac{d^2{\bf k}_{i\bot}d\rho(int_i)}{2(2\pi)^3\xi_i(1-\xi_i)},\nonumber\\
&&{\tilde \rho}_i(\xi_i)=\sum_{\sigma_i}\int\nolimits
|{\tilde \chi}(\xi_i,{\bf k}_{i\bot},\sigma_i,int_i)|^2
\frac{d^2{\bf k}_{i\bot}d\rho(int_i)}{2(2\pi)^3\xi_i(1-\xi_i)}
\label{43}
\end{eqnarray}
where $int_i$ means the internal variables for the system
$(1,...i-1,i+1,...)$ and the integration over these variables is
assumed. As follows from Eqs. (\ref{31}-\ref{33}) and (\ref{36}),
the quantity $\rho_i(\xi_i)d\xi_i$ is the probability of the
event that the light cone momentum fraction of quark $i$ is in the
interval $(\xi_i,\xi_i+d\xi_i)$.
For this reason the functions $\rho_i(\xi_i)$ are just the
PDFs considered in the FT (streaktly speaking they coincide with the
PDFs only in the light cone gauge for the gluon field
\cite{CSS1,Sterman}).

In the parton model the current operator is replaced by
the free one and Eqs. (\ref{36}) and (\ref{37}) make it possible to
explicitly calculate the structure functions of DIS.

 A standard calculation for the unpolarized structure
functions $F_1$ and $F_2$ (see e.g. ref. \cite{Coester}) gives
\begin{equation}
F_1(x_{Bj})=\frac{1}{2}\sum_{i}e_i^2 \rho_i(x_{Bj}),\quad
F_2(x_{Bj})=x_{Bj}\sum_{i} e_i^2\rho_i(x_{Bj})
\label{44}
\end{equation}
where $e_i$ is the electric charge of particle $i$. This result
shows that the Bjorken variable $x_{Bj}$ can be interpreted as the
light cone momentum fraction of the struck quark.
At the same time if the current operator is given by Eq.
(\ref{39}) we have obviously
\begin{equation}
F_1(x_{Bj})=\frac{1}{2}\sum_{i}e_i^2{\tilde \rho}_i(x_{Bj}),\quad
F_2(x_{Bj})=x_{Bj}\sum_{i} e_i^2{\tilde \rho}_i(x_{Bj})
\label{45}
\end{equation}

 Since the wave function of the moving nucleon is defined by the function
$\chi$ and not ${\tilde \chi}$ (see Eq. (\ref{36})), only the functions
$\rho_i$ are universal while the functions ${\tilde \rho}_i$ are not.
As follows from what has been said about the operator $A$,
the relations between them are rather complicated and
the functions ${\tilde \rho}_i$ are not defined uniquely.

 As noted in the preceding section, the sum rules in DIS depend on
the form of the operators ${\hat O}_i^{\mu_1...\mu_n}$. If the
matrix elements of the current operator are calculated in the
reference frames discussed above, then, as follows from Eqs. (\ref{6})
and (\ref{42}), we can use the relation
\begin{equation}
{\hat O}_i^{\mu_1...\mu_n}=A{\tilde O}_i^{\mu_1...\mu_n}A^{-1}
\label{46}
\end{equation}
Therefore the operators ${\hat O}_i^{\mu_1...\mu_n}$ in our model
indeed contains the nonperturbative part but this part is present
only in the operator $A$.

 The momentum carried by quarks is defined by sum rules for
the functions $\xi_i\rho_i(\xi_i)$ and in the general case
\begin{equation}
\int_{0}^{1} \xi_i\rho_i(\xi_i)d\xi_i \neq
\int_{0}^{1} \xi_i{\tilde \rho}_i(\xi_i)d\xi_i
\label{47}
\end{equation}
Therefore if the current operator is given by Eq. (\ref{39})
or Eq. (\ref{40}), then DIS experiments
cannot directly determine the fraction of the nucleon momentum
carried by quarks. Analogously these
experiments cannot directly determine the fraction of the nucleon
spin carried by quarks. It is also unclear how one can extract
from DIS experiments the PDFs which should be used for the description
of Drell-Yan pairs and hard $p\bar{p}$ collisions.

 As follows from the above discussion we have to choose between the
following possibilities
\begin{itemize}
\item If we want to work only with universal PDFs, we have to work
in the representation where the current operator is given by Eqs.
(\ref{39}) or (\ref{40}). Then the amplitudes of lepton-parton
interactions cannot be calculated in perturbation theory. In other
words, the condition $2_{FT})$ is satisfied but the condition
$1_{FT})$ is not.

\item On the other hand, we can exclude the $A$ dependence of the
current operator by using a proper unitary transformation. Then we
will have the situation described by Eq. (\ref{45}) when the
condition $1_{FT})$ is satisfied but the condition $2_{FT})$ is not.
\end{itemize}

 We see that if the possibility (\ref{40}) is realized in nature
then one can satisfy either $1_{FT})$ or $2_{FT})$ but not the both
conditions simultaneously.

  Since $\chi$ is obtained from ${\tilde \chi}$ by a unitary
transformation, it is clear from Eq. (\ref{43}) that
\begin{equation}
\int_{0}^{1} \rho_i(\xi_i)d\xi_i =
\int_{0}^{1} {\tilde \rho}_i(\xi_i)d\xi_i =1
\label{48}
\end{equation}
It is easy to show that for this reason the Adler, Bjorken and Gross -
Llewellyn Smith sum rules for unpolarized DIS \cite{Adl,Bjor1,GLS} are
satisfied also in the
case when the current operator is given by Eq. (\ref{40})
(according to the present theory, in leading approximation in
QCD running coupling constant these sum rules are given by the
parton model). Therefore the fact that the operators
${\hat O}_i^{\mu_1...\mu_n}$ and ${\tilde O}_i^{\mu_1...\mu_n}$
differ each other does not affect the sum rules \cite{Adl,Bjor1,GLS}.

At the same time the question arises whether other sum rules
are satisfied. In the framework of the OPE the validity of the Bjorken
sum rule for polarized DIS \cite{Bj} is a consequence of the
fact that ${\hat O}_A^{\mu}$ coincides with the axial current operator
${\hat J}_A^{\mu}$ (see Eqs. (\ref{7}) and (\ref{10})). In our
model this sum rule will be satisfied if the axial current operator in
the $\beta$ decay satisfies Eq. (\ref{40}) at $x=0$.
However such a choice is defined by low-energy physics and
is an additional assumption. The fact that the Bjorken
sum rule \cite{Bj} involves an assumption about the relation between
the current operators in DIS and in the $\beta$ decay has been pointed
out by several authors (see e.g. ref. \cite{Prep}).

\section{Discussion}
\label{S7}

 The results of the present paper give grounds to conclude that if
the operators ${\hat J}^{\mu}({\bf x})$ in the instant form and
${\hat J}^{\mu}(x^-,{\bf x}_{\bot})$ in the front one can be
correctly defined then some of them inevitably depend on the
nonperturbative part of the quark-gluon interaction which contributes
to DIS even in the Bjorken limit. Then the question arises whether
the canonical equal time or light cone commutation relations between the
components of the current operators are satisfied and whether the
deviation of these
relations from canonical ones can be investigated in perturbation
theory (as noted in ref. \cite{J}, at present the only
known way of investigating the commutation relations is the
renormalizable perturbation theory).
The main problem is whether these results can add something to the
understanding of the questions formulated in Sect. \ref{S1}.

 In Sect. \ref{S6} we have considered a model where the nonperturbative
contribution to the current operator is contained only in the
unitary operator $A$ (see Eq. (\ref{40})). If this possibility is
realized in nature then one can satisfy either $1_{FT})$ or $2_{FT})$ but
not the both conditions simultaneously.

 In Sect. \ref{S5} we argue that difficulties in substantiating Eq.
(\ref{6}) beyond perturbation theory exist not only
in the next-to-leading orders in $1/Q$ (see ref. \cite{Jaffe} and
references therein) but also in the Bjorken limit.
Our results definitely show that the usual physical arguments
based on asymptotic freedom at small distances are
insufficient to substantiate the expansion (\ref{6}) beyond
perturbation theory, and, if it
takes place, this is a consequence of deeper reasons.

 The important mathematical fact which is demonstrated by the model
considered in Sect. \ref{S6} is that {\it it is not necessary to
require the validity of Eq. (\ref{6}) in all reference frames.
In other words, it is not necessary to require for the operator
product expansion to be a true operator relation. For the validity of
the standard results about the $Q^2$ evolution of the structure
functions it is quite sufficient to require the validity of Eq. (\ref{6})
only in special reference frames.}

 If Eq. (\ref{6}) is understood in such a way then, as shown in
Sect. \ref{S6}, there exists a possibility that this expression is
valid, i.e. the property $1_{OPE})$ takes place.
The validity of Eq. (\ref{6}) is some reference frames will
automatically ensure its validity in all reference frames only if
the ${\hat O}_i^{\mu_1...\mu_n}$ are true tensor operators with
respect to {\it all} transformations of the Lorentz group. However
the arguments given in Sects. \ref{S4}-\ref{S6} show that the
problem of constructing such operators is rather complicated and
the property $2_{OPE})$ is problematic.

The model considered in Sect. \ref{S6} shows that there exists a
possibility that many results of the standard theory (including
the $Q^2$ evolution of the structure functions) take place even if
the current operator contains a nonperturbative contribution which
survives in the Bjorken limit. At the same time one has to take
into account that, although the present theory has many impressive
successes in describing experimental data, several problems remain.
For example,
as noted in the recent review paper \cite{Sterman},
"...On the other hand, these analyses also are calling into
question, for the first time, the ultimate consistency of the
existing theoretical framework with all existing experimental
measurements! (This can be regarded as testimony to the progress
made in both theory and experiment - considering the fact that
contradictions come with precision, and they are a necessary
condition for discovery overlooked shortcomings and/or
harbingers of new physics.)". The existence of difficulties in
the present approach has been also demonstrated in the recent
experiments at Tevatron and HERA \cite{HERA}. It is interesting to
note that the
most serious difficulties appear at very large values of the
momentum transfer, i.e. when according to the usual philosophy
there should be no doubt about the validity of the standard approach.

 The possibility that the nonperturbative part of the current
operator is important even in the Bjorken limit was considered
by several authors (see e.g. ref. \cite{Gl} and
references therein). Our consideration differs from that of
those authors since we
tackle the problem of nonperturbative effects by considering the
commutation relations between the current operator and the
representation generators of the Poincare group.

 In view of the above discussion it is very important to
know what is the contribution of the nonperturbative effects to the
current operator but at the present stage of QCD this contribution
cannot be calculated. It is also important to
know which experiments can shed light on our understanding
of the structure of the current operator.
In ref. \cite{NPA} we argue that deuteron DIS at large $Q$ and
$x_{Bj}\leq 0.01$ is such an experiment.

\begin{center} {\bf Acknowledgments} \end{center}
\begin{sloppypar}
 The author is grateful to S.J.Brodsky, F.Coester,
A.B.Kaidalov, L.A.Kondratyuk, B.Z.Kopeliovich,
L.N.Lipatov, M.P.Locher, S.Mikhailov, N.N.Nikolaev, V.A.Novikov,
E.Pace, H.C.Pauli, R.Petronzio, A.V.Radyushkin, G.Salme,
O.Yu.Shevchenko, I.L.Solovtsov and H.J.Weber
for valuable discussions. This work was
supported in part by grant No. 96-02-16126a from the Russian
Foundation for Fundamental Research.
\end{sloppypar}

\end{document}